\documentclass[
journal = aaembp,
manuscript = article,
layout = twocolumn
]{achemso}

\usepackage{siunitx}
\usepackage{graphicx}
\usepackage{amsmath, amssymb}

\newcommand{\tfg}[1]{\textsubscript{\protect\raisebox{-1pt}{#1}}}

\newcommand{\mtfg}[1]{_{\mathrm{#1}}}			
\newcommand{\mhog}[1]{^{\mathrm{#1}}}

\newcommand{\mhtg}[2]{\mhog{#2}\mtfg{#1}}
\newcommand{\diff}{\mathrm{d}}
\newcommand{\iu}{\mathrm{i}}

\SectionNumbersOn
\AbstractOn

\title[Ni vacancy in NiO]{The nickel vacancy acceptor in NiO: doping beyond thermodynamic equilibrium}
\author{Robert Karsthof}
\affiliation{Centre for Materials Science and Nanotechnology, Universitetet i Oslo, Gaustadalléen 23A, 0373 Oslo, Norway}
\alsoaffiliation{Felix Bloch Institute for Solid State Physics, Universit\"{a}t Leipzig, Linn\'{e}str. 5, 04103 Leipzig, Germany}
\email{r.m.karsthof@smn.uio.no}

\author{Arthur Markus Anton}
\affiliation{Department of Physics and Astonomy, The University of Sheffield, Hicks Building, Hounsfiled Road, Sheffield S3 7RH, United Kingdom}
\alsoaffiliation{Peter Debye Institute for Soft Matter Physics, Universit\"{a}t Leipzig, Linn\'{e}str. 5, 04103 Leipzig, Germany}

\author{Friedrich Kremer}
\affiliation{Peter Debye Institute for Soft Matter Physics, Universit\"{a}t Leipzig, Linn\'{e}str. 5, 04103 Leipzig, Germany}

	\author{Marius Grundmann}
\affiliation{Felix Bloch Institute for Solid State Physics, Universit\"{a}t Leipzig, Linn\'{e}str. 5, 04103 Leipzig, Germany}

\begin{document}
	
\begin{abstract}
	This work reports on temperature-induced out-diffusion and concentration decay of the prominent intrinsic point defect V\tfg{Ni} (nickel vacancy) in the wide-gap $p$-type semiconductor nickel oxide (NiO). V\tfg{Ni} can easily be introduced into NiO thin films by offering high oxygen partial pressures during film growth, rendering nonstoichiometric semiconducting structures. However, exposure to lower oxygen supply after growth, e.g. in a standard atmosphere,  usually leads to a gradual decrease of film conductivity, because the vacancy concentration equilibrates. In this study, we observe this process \textit{in situ} by performing temperature-dependent measurements of the electrical conductivity on a room temperature-grown NiO film. At a temperature of \SI{420}{\kelvin} under exclusion of oxygen, the doping level decreases by a factor of 8 while the associated room temperature dc conductivity drops by six orders of magnitude. At the same time, out-diffusion of the mobile V\tfg{Ni} species can be indirectly observed through the occurrence of electrode polarization characteristics. 
\end{abstract}

\section{Introduction}
	
Nickel oxide (NiO) is a $p$-type wide-gap semiconducting oxide with promising properties for various device applications, most notably as hole-transporting electrode in solar cells \cite{He.1999,Irwin.2008,Park.2010,Jeng2014}. For these devices, NiO is typically doped intrinsically by an excess supply of reactive oxygen during film growth, which yields a large concentration of Ni vacancy defects (V\tfg{Ni}). Because V\tfg{Ni} is a double acceptor in NiO, this is the prominent method to obtain NiO thin films with semiconducting properties, apart from extrinsic doping with Li. In a recent publication \cite{Karsthof2019}, we investigated the electronic transport properties of intrinsically doped NiO thin films and ascertain that polarons of intermediate size (about two lattice constants), strongly bound to acceptors, are the dominant charge carriers in these films. Electrical current is carried by hopping of these carriers exclusively among V\tfg{Ni} sites without excitation to any extended states, i.e. band conduction is negligible. At temperatures above about half the Debye temperature $\theta\mtfg{D}$ ($\approx \SI{200}{\kelvin}$ for NiO \cite{Allen1954}) this is an over-the-barrier hopping process with an activation energy determined by the inter-site separation. At temperatures lower than $\theta\mtfg{D}/2$ the activation energy is of the order of the width of the density of states. This implies that structural and electronic disorder play a central role for the temperature dependence of the electric conductivity of intrinsically doped NiO samples. 

The degree of film nonstoichiometry is determined by the reactivity of the oxygen supply (i.e., ionized vs. neutral) during film growth. High pressures and particle velocities develop at the laser ablation site on the target, leading to a high degree of ionization of the oxygen background gas. This favors a high oxidation of the growing film. When exposed to ambient atmosphere, samples prepared in this way can be expected to be long-term unstable with regard to the Ni vacancy concentration. As a result, their conductivity is expected to decay with time, which is accelerated at elevated temperatures. This publication reports on the first, partially \textit{in situ}, experimental observation of this process. We believe the derived results are of importance for design rules of devices containing an intrinsically doped NiO layer, as in the case of hole-transport layers in solar cells for instance.

\section{Experimental details}
	
The NiO layers described here were fabricated by pulsed laser deposition (PLD), using a KrF excimer laser (wavelength \SI{248}{\nano\meter}, energy per pulse \SI{650}{\milli\joule}, repetition rate \SI{10}{hertz}) from a ceramic NiO target (Alfa Aesar, purity \SI{99.998}{\percent}). During deposition, the O\tfg{2} partial pressure was kept at \SI{0.1}{\milli\bar}, and no intentional substrate heating was employed (room temperature growth). Two types of NiO thin film samples were fabricated. The first one consisted of NiO thin films deposited on a metallic Pt electrode on top of corundum substrates. These samples were used to study the electronic transport. The second type was NiO deposited on top of commercial fluorine-doped tin oxide (FTO)-covered glass substrates. This combination was employed in order to induce a space charge region in the NiO layer, which enabled the use of space charge spectroscopy to determine the doping level of the NiO. For both types, the NiO layers were capped by \SI{20}{\nano\meter} thick Pt layers. The deposition of NiO and Pt capping through a priorly structured standard UV photoresist allowed their patterning by the lift-off technique into pillars with circular cross sections (diameters between \SI{250}{\micro\meter} and \SI{800}{\micro\meter}). All Pt layers were fabricated by dc magnetron sputtering.

Current-voltage characterization was done using a semi-automatic S\"{U}SS wafer prober with tungsten needles and an AGILENT 4155C Precision Semiconductor Parameter Analyzer. The results of these measurements were used to select individual contacts for further characterization. 

For capacitance-voltage measurements, the samples were mounted onto transistor sockets, and the selected contacts were wire-bonded using Au wires and silver epoxy resin. For broadband dielectric spectroscopy (BDS) measurements, a similar method was applied, using a home-built sample holder instead of a transistor socket. The silver epoxy resin was hardened in a dry cabinet at \SI{90}{\degreeCelsius} for \SI{30}{\minute}.

BDS measurements were carried out in a temperature and frequency range between \SI{300}{\kelvin} and \SI{420}{\kelvin} and between \SI{e-2}{\hertz} and \SI{e7}{\hertz}, respectively, by means of a NOVOCONTROL Technology high-resolution $\alpha$-analyzer combined with a Quatro temperature controller (absolute thermal stability $\le \SI{1}{\kelvin}$). 

X-ray diffraction was performed using a Philips X'Pert diffractometer and the Cu $K\alpha$ line ($\lambda = \SI{1.5406}{\angstrom}$).

\section{Results}

\begin{figure}
	\centering
	\includegraphics[width=\columnwidth]{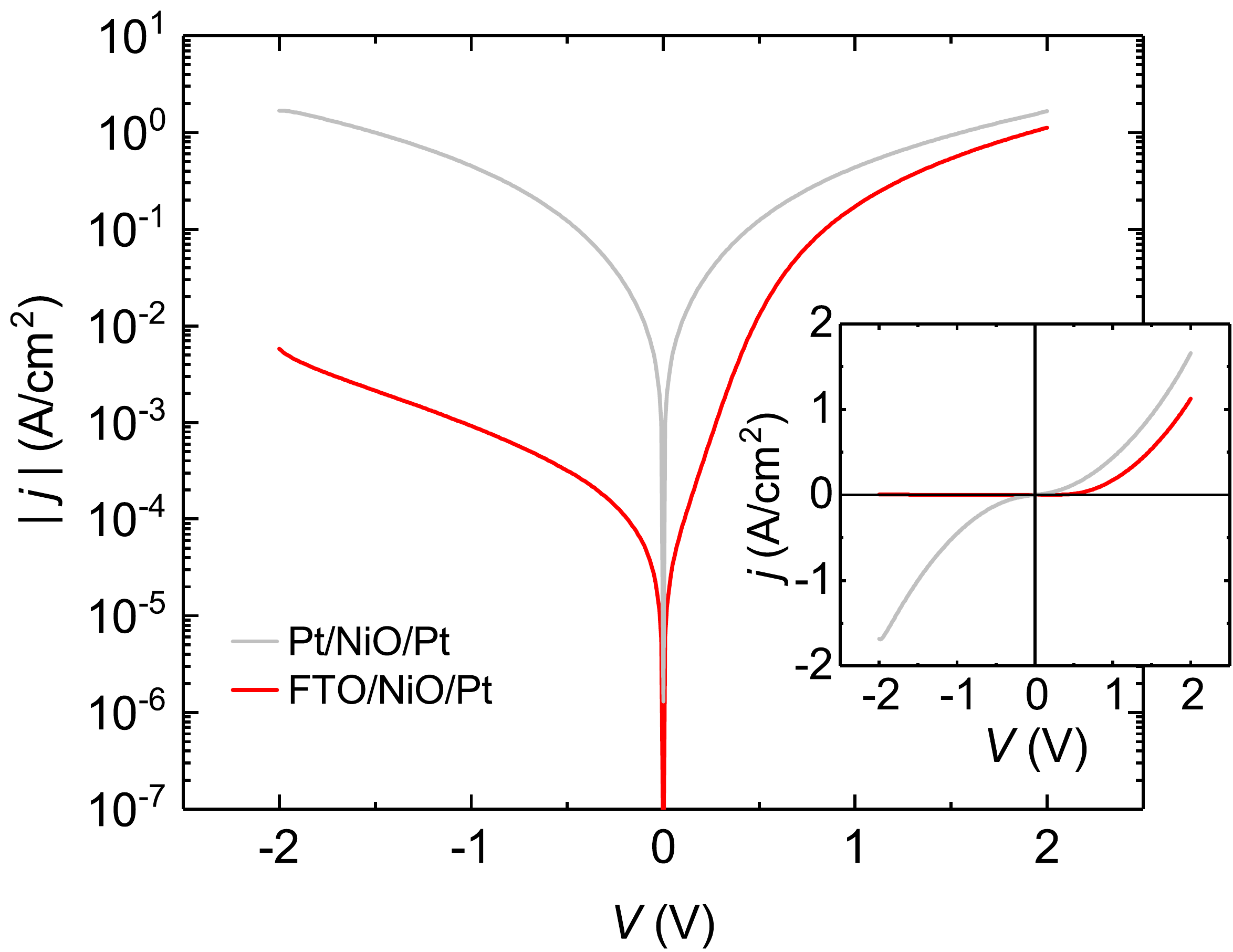}
	\caption{Current density-voltage relations of selected Pt/NiO/Pt and FTO/NiO/Pt structures in semilogarithmic and linear (inset) scale.}
	\label{fig:NiO-Pt+FTO_jV}
\end{figure}

The current density-voltage relations of the two structures shown in Fig.~\ref{fig:NiO-Pt+FTO_jV} demonstrate that the NiO layer on FTO exhibits a current rectification of approximately two orders of magnitude at $\pm \SI{2}{\volt}$, whereas the characteristics of the layer sandwiched between Pt electrodes show symmetric behavior with respect to voltage. The latter is therefore appropriate for the investigation of electrical transport, as previously done in Ref.~\cite{Karsthof2019}. It shall be noted that the transport is non-ohmic, i.e. the differential conductivity increases with increasing voltage (see inset of Fig.~\ref{fig:NiO-Pt+FTO_jV}). The same effect is observed in the forward region of the FTO/NiO diode (for voltage above $\approx \SI{1}{\volt}$). This behavior can be attributed to the specific dc transport mechanism of NiO which has similarities with space charge-limited conduction (SCLC) \cite{Rose.1955,Seo.2005,Chang.2006}.  

There are only few reports of rectifying contacts comprising NiO as the carrier-depleted semiconductor. It has been shown that the low work function metals Al and Ti can be used to fabricate Schottky contacts on NiO \cite{Seo.2005}, and the transparent conductive oxides tin-doped indium oxide (ITO) \cite{Chang.2006} and fluorine-doped tin oxide (FTO) \cite{Thimsen2012} have been proven to be appropriate $n$-type partners for rectifying $pn\mhog{+}$ heterostructures. A depletion region in a semiconductor enables the application of space charge spectroscopy techniques to investigate defect properties, such as the doping level and its spatial distribution. This is especially useful in the case of NiO because the determination of charge carrier concentration by Hall effect measurements is difficult due to extraordinarily low charge carrier mobility ($\mu\mtfg{p} \ll \SI{0.1}{\square\centi\meter\per\volt\per\second}$).

\begin{figure}
	\centering
	\includegraphics[width=\columnwidth]{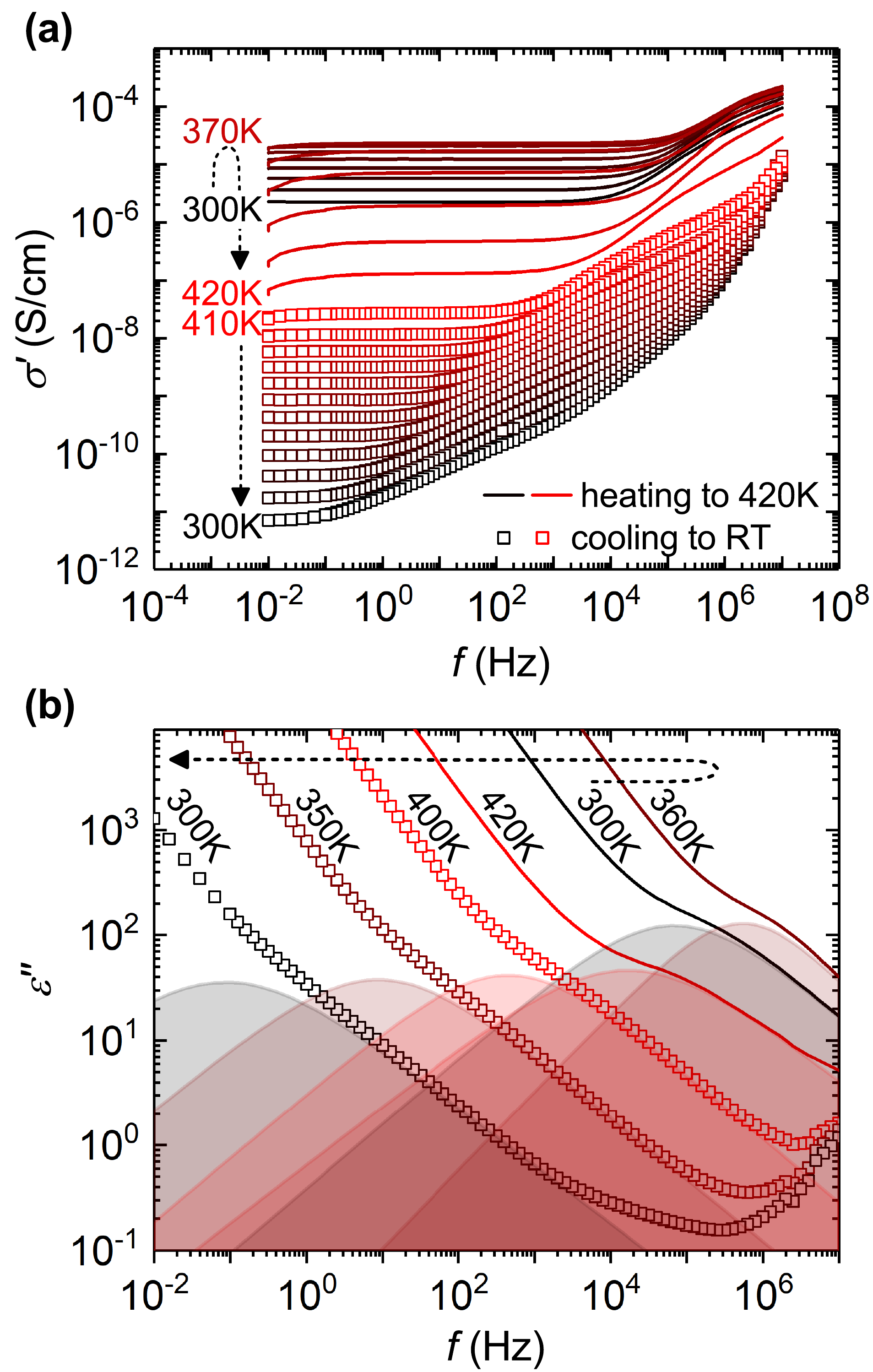}
	\caption{(a) real part $\sigma'$ of the complex conductivity function $\sigma\mhog{*} = \sigma' + \iu \sigma''$, measured by BDS in a temperature cycle between room temperature and \SI{420}{\kelvin}. (b) selected temperatures from the same dataset in the dielectric loss representation, including fits to the MWS polarization acc. to (\ref{eq:cole-cole}).}
	\label{fig:BDS_anneal_sigma'+eps''}
\end{figure}

In a recent publication, we reported on the electrical transport properties of NiO below room temperature. In the present work, we address the same issue for elevated temperatures up to \SI{420}{\kelvin}. Fig.~\ref{fig:BDS_anneal_sigma'+eps''}a shows the real part $\sigma'$ of the complex conductivity ${\sigma\mhog{*} = \sigma' + \iu \sigma''}$ of the Pt/NiO/Pt sample, recorded in a temperature cycle between \SI{300}{\kelvin} and \SI{420}{\kelvin}. In general, the frequency dependence of $\sigma'$ of the films can be separated into two different regions. At \SI{300}{\kelvin} at the beginning of the cycle, $\sigma'$ is independent of the frequency of the electric field, $f$, up to \SI{e3}{\hertz}. This is interpreted here as the dc conductivity limit of the electronic conduction process, $\sigma\mtfg{dc}$. With increasing $f$, $\sigma'$ begins to exhibit dispersion, rising slowly in the form of a broadened step. In our recent paper \cite{Karsthof2019}, we showed that this is the signature of a dielectric polarization process connected to the spatial inhomogeneity of $\sigma\mhog{*}$ (Maxwell-Wagner-Sillars (MWS) polarization \cite{Kremer}) which is caused by the large degree of disorder present in room temperature-deposited NiO. 

When the complex dielectric function $\varepsilon\mhog{*}$ is calculated from $\sigma\mhog{*}$ according to

\begin{equation}
		\varepsilon\mhog{*} = \frac{1}{2\pi\iu f \varepsilon_0} \sigma\mhog{*},
\end{equation} 

relaxation and polarization processes are visible as peaks in the imaginary part of $\varepsilon\mhog{*}$, $\varepsilon'' = \frac{1}{\omega\varepsilon_0} \sigma'$. These peaks can usually be fitted with a Debye-type function or modifications thereof; in the present case, we used the Cole-Cole relation

\begin{equation}
 \varepsilon\mhog{*} = \varepsilon\mtfg{\infty} + \frac{\varepsilon\mtfg{s}-\varepsilon\mtfg{\infty}}{1+\left(\iu \omega \tau\mtfg{rel}\right)\mhog{1-b}} 
\end{equation}

with $\varepsilon\mtfg{\infty}$ and $\varepsilon\mtfg{s}$ the high-frequency and static value of the relative dielectric permittivity, $\omega = 2\pi f$, $\tau\mtfg{rel}$ the relaxation time constant, and $b \in [0,1)$ a dimensionless broadening parameter, respectively. Taking additional account of the dc conductivity, the following relation was used to fit the $\varepsilon''$ data:

\begin{equation}
	\begin{split}
		\varepsilon''(&\omega) = \frac{\sigma\mtfg{dc}}{\omega \varepsilon_0} \\
		&+ \frac{\Delta \varepsilon}{2} \frac{\cos b\pi /2}{\cosh \left[(1-b)\ln\left(\omega\tau\mtfg{rel}\right) \right] + \sin b\pi /2},
	\end{split}
	\label{eq:cole-cole}
\end{equation}

where $\sigma\mtfg{dc}$ is the dc conductivity taken as the value of $\sigma'$ at the low-frequency plateau and ${\Delta\varepsilon = \varepsilon\mtfg{s} - \varepsilon\mtfg{\infty}}$. The imaginary part of the dielectric function, $\varepsilon''$, and the fits to the MWS polarization process according to (\ref{eq:cole-cole}) (without the dc contribution), is shown in Fig.~\ref{fig:BDS_anneal_sigma'+eps''}b for selected temperatures. 

A simplifying model of interfacial polarization processes due to a spatially varying dielectric function considers spherical filler particles (volume fraction $\varphi\mtfg{f}$) with conductivity $\sigma'\mtfg{f}$ embedded in a matrix with lower conductivity $\sigma'\mtfg{m} \ll \sigma'\mtfg{f}$. Both media are assumed to have the same real part of the dielectric function $\varepsilon'$. It can be shown \cite{Kremer} that in this case a Debye-like relaxation peak, representing Maxwell-Wagner-Sillars polarization, occurs, with 

\begin{align}
	\Delta\varepsilon\mtfg{MWS} &\approx 3\varepsilon'\left(\frac{\sigma'\mtfg{f}}{\sigma'\mtfg{m}}\right)^2\frac{\varphi\mtfg{f}\left(1-\varphi\mtfg{f}\right)}{\left(2+\varphi\tfg{f}\right)^2}, \; \text{and} \label{eq:DeltaEpsMWS} \\
	\tau\mtfg{rel,MWS} &\approx 3\varepsilon_0 \varepsilon' \frac{1}{\left(1-\varphi\mtfg{f}\right)\sigma'\mtfg{f}}. \label{eq:tauMWS}
\end{align} 

Therefore, depending on the 'conductivity contrast' between filler and matrix, $\Delta\varepsilon$ can attain larger values than commonly considered for molecular relaxation processes (typically $\Delta\varepsilon \ll 1$). Moreover, (\ref{eq:tauMWS}) demonstrates that the relaxation rate $\omega\mtfg{rel,MWS} = \frac{2\pi}{\tau\mtfg{rel,MWS}}$ is directly proportional to the conductivity of the filler, implying that it roughly exhibits the same activation energy.\\

\begin{figure}
	\centering
	\includegraphics[width=\columnwidth]{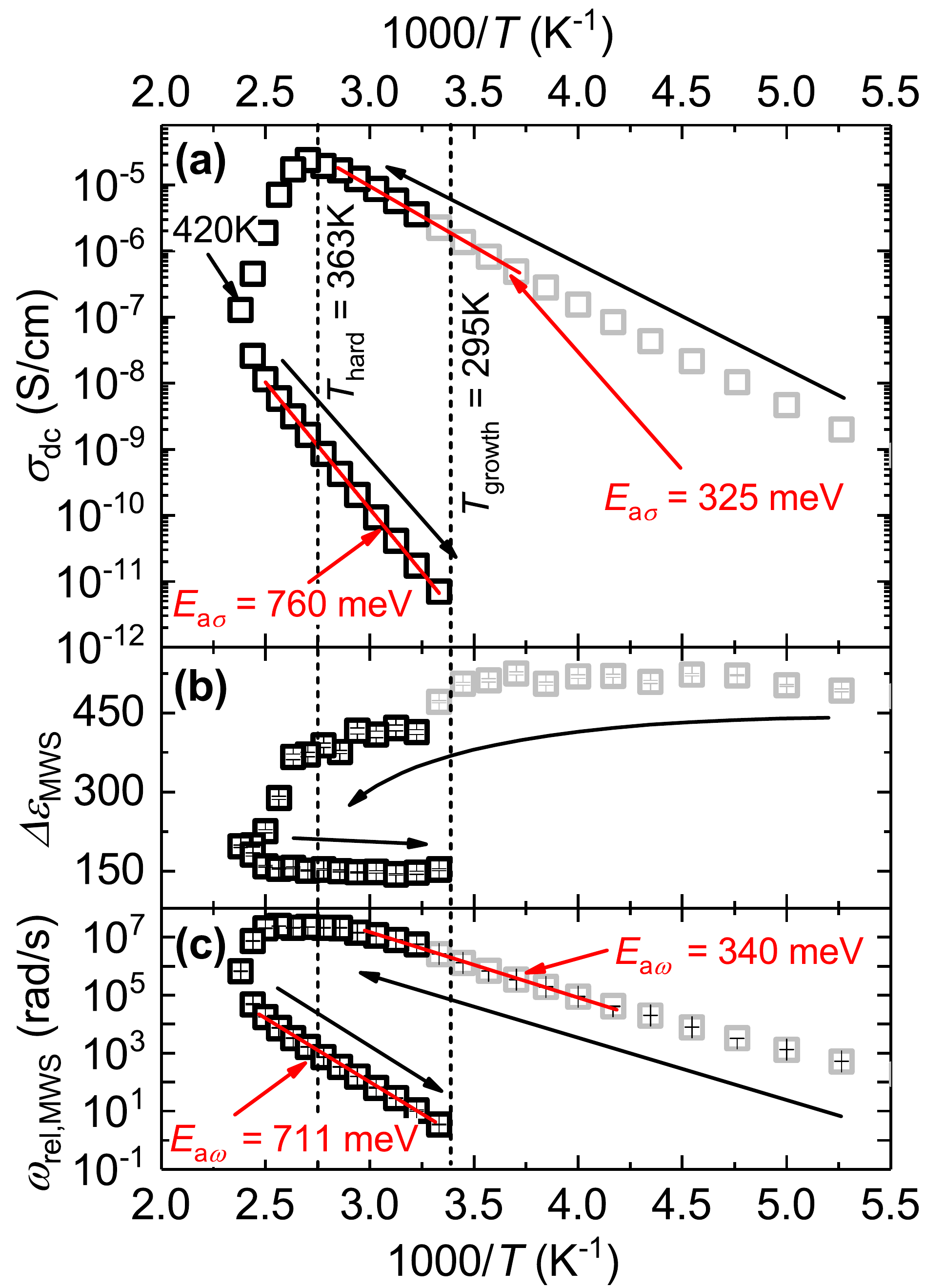}
	\caption{(a) dc conductivity, (b): relaxation strength and (c): rate of the MWS polarization mode for temperatures above \SI{300}{\kelvin} (black symbols). Representative data points from measurements below \SI{300}{\kelvin} are shown in addition (gray symbols) \cite{Karsthof2019}.}
	\label{fig:BDS_sigmaDC+DeltaEps+wMWS_hot}
\end{figure}

Because the film was grown at room temperature and the measurement is done at elevated temperatures, the presented experiment can be considered an \textit{in situ} annealing experiment conducted under oxygen-poor conditions (pure N\tfg{2} atmosphere). At temperatures above about \SI{370}{\kelvin} three annealing-induced features appear simultaneously: (i) The frequency-independent contribution $\sigma\mtfg{dc}$ drops, even though the temperature is increased to \SI{420}{\kelvin}. During subsequent cooling to room temperature, $\sigma\mtfg{dc}$ decreases further, as expected due to the decreasing temperature, resulting in a final reduction of six orders of magnitude as compared to the initial (room temperature) value (see Fig.~\ref{fig:BDS_sigmaDC+DeltaEps+wMWS_hot}, top panel). (ii) Electrode polarization sets in, causing a drop in $\sigma'$ at very low frequencies, (iii) the MWS polarization process strongly decreases in both relaxation strength $\Delta\varepsilon\mtfg{MWS}$ and relaxation rate ${\omega\mtfg{MWS}}$ (Fig.~\ref{fig:BDS_sigmaDC+DeltaEps+wMWS_hot}, center and bottom panel).

We attribute all these observations to the thermally induced out-diffusion of Ni vacancies. Because the initial concentration of V\tfg{Ni} is much higher than the thermodynamically adequate level, the film has the tendency to equilibrate this concentration at any given temperature, especially at $T>T\mtfg{growth}$. This is achieved by the diffusion of V\tfg{Ni} to the sample surface where they decay by locally dissolving the crystal\footnote{There are experiments that have directly detected the reversion of this process (e.g., the creation of Ni vacancies at elevated temperatures through the equilibration in O\tfg{2} gas) by thermogravimetric methods, for example by Mitoff \cite{Mitoff.1961}, and Osburn and Vest \cite{Osburn.1971a}.} according to the reaction

\begin{equation}
2\, \text{V}\mhtg{\text{Ni}}{+2-c} + 2\,\text{O}\mhtg{\text{O}}{-2} \longrightarrow \text{O\tfg{2}(g)} + 4\, e\mhog{-} + (4-2c)\,h\mhog{+}.
\label{eq:VNi_outdiffusion}
\end{equation}

This renders the film interior less conductive. Eqn.~(\ref{eq:VNi_outdiffusion}) also includes the formal charges that are attributed to the constituents. Ideally, each O site carries a formal $-2$ charge while that of each V\tfg{Ni} acceptor is $+2$. Due to the presence of compensating donor states ("hole killers"), however, a fraction $0<c<1$ of the vacancies is ionized. The reaction therefore produces a net charge of $-2c\cdot e$ per destroyed vacancy site. Since the sample surface consists mostly of the film/electrode interface (film thickness $\ll$ contact diameter), these charges temporarily accumulate in its vicinity and thereby partially screen the interior of the sample from the applied AC electric field, before they migrate into the electrode. This leads to an apparently lower electric conductivity of the sample (electrode polarization \cite{Kremer,Serghei.2009,Ishai.2013}, EP). Because this effect is in principle comparable to ionic conduction, the characteristic frequency of an EP process is determined by the ion diffusivity and is therefore typically low; here, EP can be seen to dominate $\sigma'$ below \SI{1}{\hertz} (Fig.~\ref{fig:BDS_sigmaDC+DeltaEps+wMWS_hot}a). 

On the basis of the conductivity model developed in our recent work \cite{Karsthof2019}, a decrease of the acceptor density is expected to be accompanied by an increase in the activation energy of $\sigma\mtfg{dc}$. This is the direct result of the increased inter-acceptor distance, due to the V\tfg{Ni} decay, and observed here: from $E\mtfg{a,\sigma} = \SI{325}{\milli\electronvolt}$ before to \SI{760}{\milli\electronvolt} after the high-temperature BDS measurement. The decrease of $\Delta\varepsilon\mtfg{MWS}$ and $\omega\mtfg{rel,MWS}$ of the MWS polarization process as a consequence of annealing is also in accordance with the model, because both parameters are affected by the conductivity of the film (Eqns.~(\ref{eq:DeltaEpsMWS}) and (\ref{eq:tauMWS})). In particular, $\omega\mtfg{rel,MWS}$ exhibits almost the same activation energy (\SI{340}{\milli\electronvolt} before and \SI{711}{\milli\electronvolt} after annealing) as $\sigma\tfg{dc}$.

In order to relate the decrease of conductivity to a lower V\tfg{Ni} density, a FTO/NiO sample was used as a reference. Due to the occurrence of charge depletion in the NiO layer, $CV$ measurements could be conducted on this sample. It was subjected to a thermal treatment at \SI{420}{\kelvin} in N\tfg{2} atmosphere for two hours to mimic the conditions during the high-temperature BDS measurement. 

\begin{figure}
	\centering
	\includegraphics[width=\columnwidth]{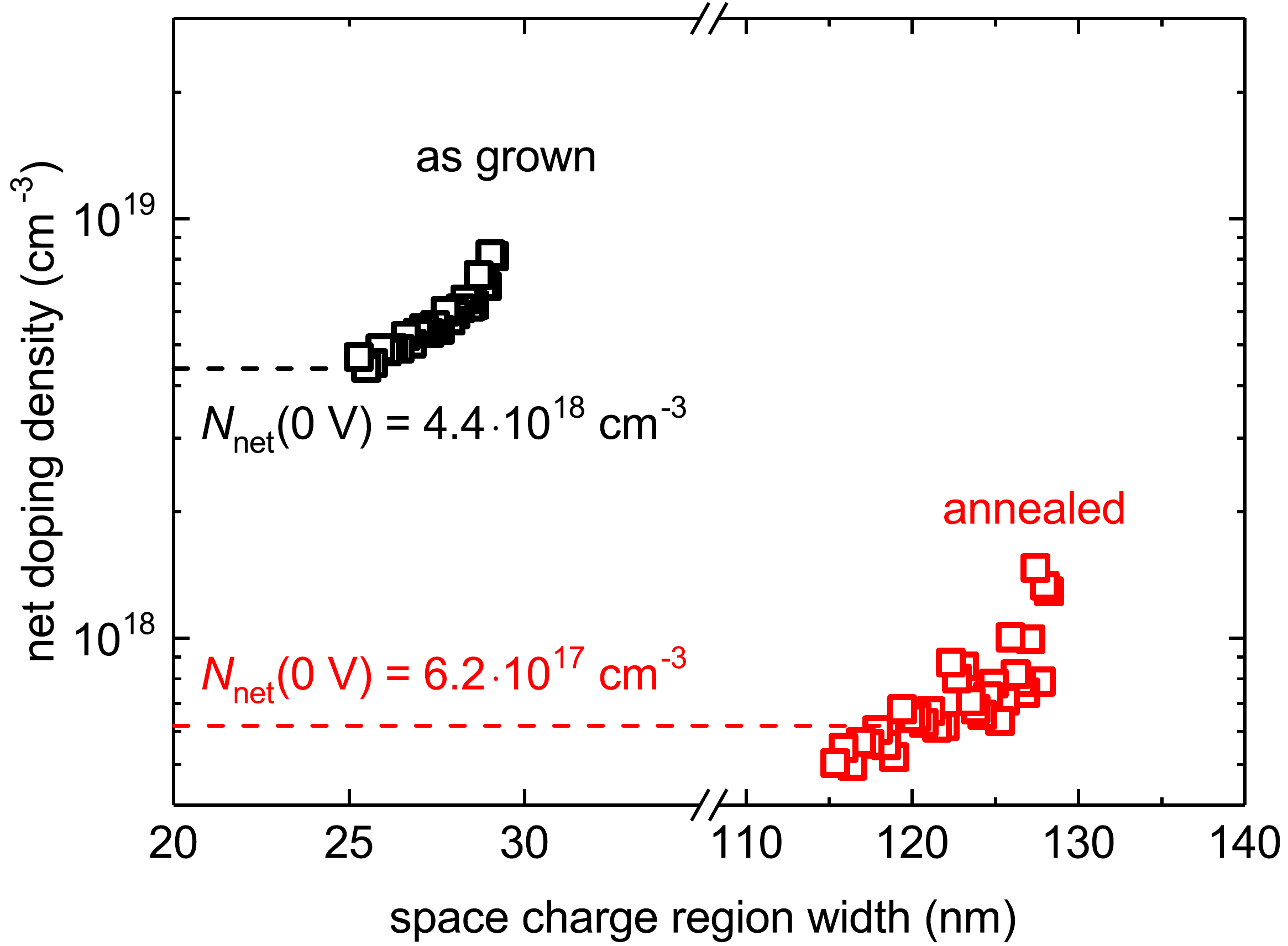}
	\caption{Doping profiles of a selected NiO/FTO contact before and after annealing at \SI{420}{\degreeCelsius} in N\tfg{2}.}
	\label{fig:NiO-FTO_doping_profiles}
\end{figure}

The doping profiles can be extracted from the $C$-$V$ measurements via

\begin{align}
	N\mtfg{net} &= \frac{2}{e\varepsilon\mtfg{s}\varepsilon_0 A^2}\left(\frac{\diff}{\diff V} C\mhog{-2}\right)\mhog{-1}, \nonumber \\
	w&= \frac{\varepsilon\mtfg{s}\varepsilon_0 A}{C}.	
\end{align}

The averaged net doping obtained by this method decreases from initially \SI{4.4e18}{\per\cubic\centi\meter} to \SI{6.2e17}{\per\cubic\centi\meter}. Because the doping level is mainly determined by the dominant acceptor V\tfg{Ni}, this result confirms the out-diffusion of the V\tfg{Ni} acceptors and corroborates the distinct decrease of the film conductivity. \\

X-ray diffraction (XRD) measurements were carried out on a Pt/NiO reference sample before and after an annealing step (\SI{420}{\kelvin} in N\tfg{2} atmosphere for two hours). The position and broadening of the NiO (111) reflex remains unchanged by this procedure (see Fig. 1, Supporting Information), which suggests that a temperature of \SI{420}{\kelvin} does not induce significant structural changes to room temperature-grown NiO films. Therefore, we conclude that the observed annealing-induced changes of electrical transport properties are entirely due to the removal of V\tfg{Ni} defects.

\begin{figure*}
	\centering
	\includegraphics[width=2\columnwidth]{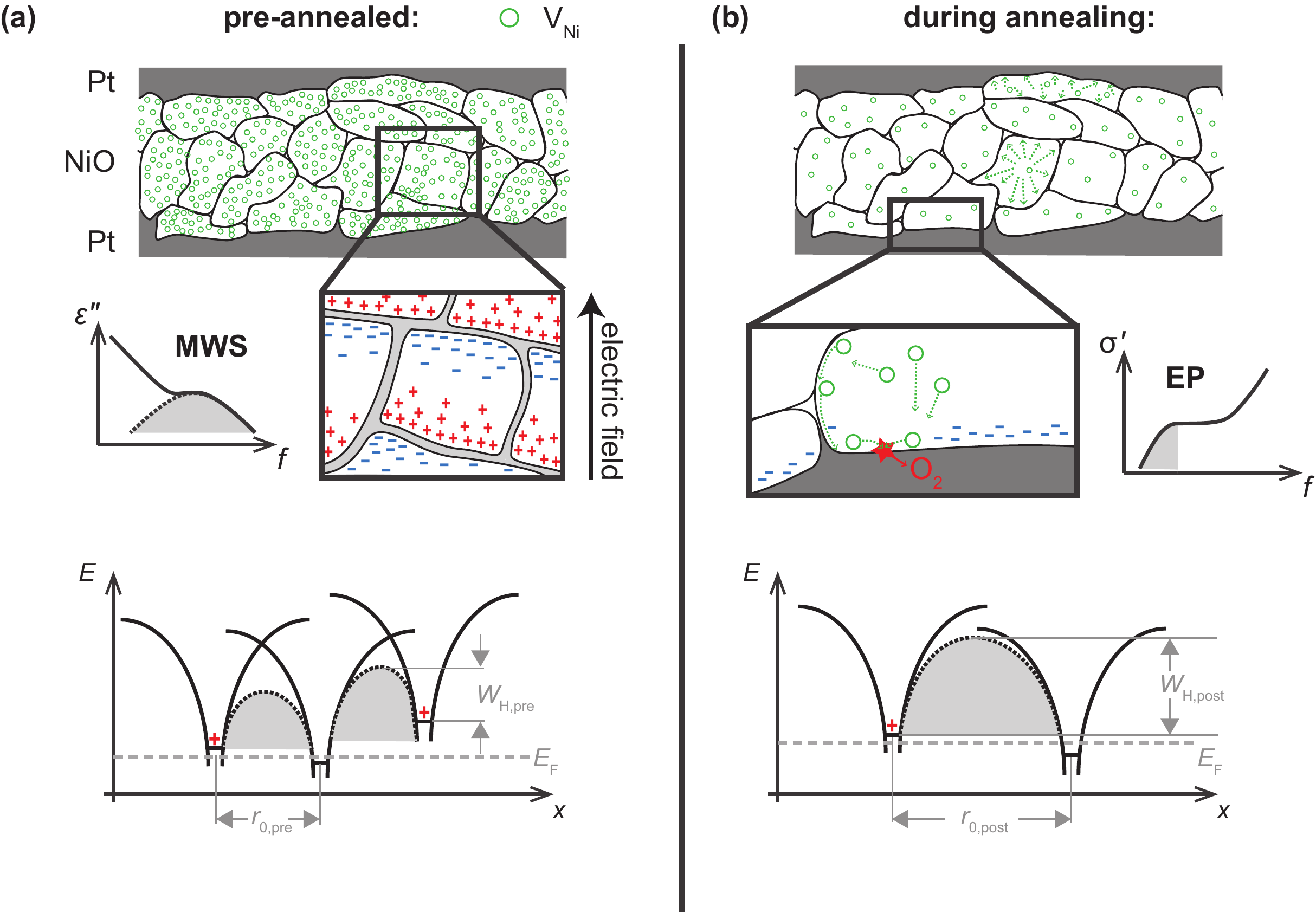}
	\caption{Annealing-induced effects of V\tfg{Ni} out-diffusion: (a) schematic cross-sectional view of the film before annealing with high density of V\tfg{Ni}, (b) polarization of mobile charges within grains leads to dielectric loss peak (Maxwell-Wagner-Sillars polarization), (c) inter-acceptor barriers for hopping transport are low. During annealing, the V\tfg{Ni} concentration decreases due to (d) out-diffusion to and (e) oxygen release at the film/ electrode interface, producing electrode polarization. Lower V\tfg{Ni} concentration results in enhanced hopping barriers between adjacent acceptors (f).}
	\label{fig:scheme_NiO}	
\end{figure*}

The observed processes are summarized in Fig.~\ref{fig:scheme_NiO} and as follows. The growth of NiO films by PLD takes place under conditions strongly deviating from thermodynamic equilibrium. This leads to a concentration of defects in the growing film that is considerably higher than in thermodynamic equilibrium, which holds in particular true for V\tfg{Ni} as it is the dominant intrinsic point defect  (Fig.~\ref{fig:scheme_NiO}a). Charge transport is facilitated through hopping conduction over initially low barriers between neighboring vacancy sites (Fig.~\ref{fig:scheme_NiO}c). On the other hand, the structural inhomogeneity gives rise to the MWS polarization of conductive grains embedded in a semi-insulating matrix (grain boundaries) which produces the according spectral relaxation process shown in Fig.~\ref{fig:scheme_NiO}b. 

At elevated temperatures, the V\tfg{Ni} defects become mobile. They diffuse via consecutive hopping processes of Ni ions to vacancies. The activation energy for this migration has been determined by several authors \cite{Shim1957,Atkinson1979} to be in the range of \SI{1.92}{\electronvolt} to \SI{2.56}{\electronvolt}. A large fraction of diffusing vacancies reaches the film/electrode interfaces, where they dissolve. At room temperature, the diffusion is not efficient, leading to a rather slow decay of the vacancy concentration. Because the films have been kept at room temperature before the BDS measurement for several days, however, some equilibration may already have taken place. Increasing the temperature to above \SI{300}{\kelvin} significantly enhances the decay rate: $\omega\mtfg{rel,MWS}$ rises slower than below room temperature (see also Fig. 2, Supporting Information). In addition, $\Delta\varepsilon\mtfg{MWS}$ clearly drops from $\approx 520$ to $\approx 350$ between \SI{300}{\kelvin} and \SI{370}{\kelvin}. However, the most pronounced acceleration of the V\tfg{Ni} decay process can be seen above \SI{360}{\kelvin}, because this was the temperature at which the sample was kept during hardening the silver epoxy. Furthermore, as soon as the temperature is lowered again, the diffusion of V\tfg{Ni} slows down dramatically. When the sample is exposed to a second temperature cycle (see Fig. 3, Supporting Information), the out-diffusion again increases above \SI{300}{\kelvin}, however, to a much lesser extent than during the first run.\\
This out-diffusion process is important to monitor when designing devices that are based on semiconducting, Ni-deficient NiO. In many technologically relevant cases, reactively sputtered NiO films are employed, where the deviation from stoichiometry is even more pronounced. When such films are used in devices where operation leads to heating, the device performance can be critically impaired by the decay of the V\tfg{Ni} acceptor.

\section{Conclusion}

Room temperature-grown NiO thin films were investigated by broad-band dielectric spectroscopy (BDS) measurements between \SI{300}{\kelvin} and \SI{420}{\kelvin}, allowing an \textit{in situ} observation of annealing-induced effects on the frequency- and temperature-dependent conductivity $\sigma'$. It was shown that out-diffusion of the dominant intrinsic acceptor V\tfg{Ni} (Ni vacancy) leads to a lower net doping level of the film (about a factor of 8), as determined by capacitance-voltage profiling. This, in turn, decreases the film conductivity at room temperatures by six orders of magnitude. The existence of mobile ionic species to which the charges are bound is supported by the observation of electrode polarization effects that accompany the drop of $\sigma\mtfg{dc}$. Simultaneously, strength and relaxation rate of a polarization process, that dominates the dispersion of the dielectric function, are strongly reduced. This is in accordance with our previous assignment of this relaxation process as a phenomenon induced by spatially varying $\sigma'$, also known as Maxwell-Wagner-Sillars polarization.

\begin{acknowledgement}
	This work was funded by the Deutsche Forschungsgemeinschaft (DFG, German Research Foundation) – project number 31047526, SFB762, project B06 and “SFB/TRR 102: Polymers under Multiple Constraints” (Project No. B08). In addition, A.M.A. is thankful financial support from the DFG project AN 1523/1-1.
\end{acknowledgement}


\bibliography{ref}

\providecommand{\latin}[1]{#1}
\makeatletter
\providecommand{\doi}
  {\begingroup\let\do\@makeother\dospecials
  \catcode`\{=1 \catcode`\}=2 \doi@aux}
\providecommand{\doi@aux}[1]{\endgroup\texttt{#1}}
\makeatother
\providecommand*\mcitethebibliography{\thebibliography}
\csname @ifundefined\endcsname{endmcitethebibliography}
  {\let\endmcitethebibliography\endthebibliography}{}
\begin{mcitethebibliography}{18}
\providecommand*\natexlab[1]{#1}
\providecommand*\mciteSetBstSublistMode[1]{}
\providecommand*\mciteSetBstMaxWidthForm[2]{}
\providecommand*\mciteBstWouldAddEndPuncttrue
  {\def\EndOfBibitem{\unskip.}}
\providecommand*\mciteBstWouldAddEndPunctfalse
  {\let\EndOfBibitem\relax}
\providecommand*\mciteSetBstMidEndSepPunct[3]{}
\providecommand*\mciteSetBstSublistLabelBeginEnd[3]{}
\providecommand*\EndOfBibitem{}
\mciteSetBstSublistMode{f}
\mciteSetBstMaxWidthForm{subitem}{(\alph{mcitesubitemcount})}
\mciteSetBstSublistLabelBeginEnd
  {\mcitemaxwidthsubitemform\space}
  {\relax}
  {\relax}

\bibitem[He \latin{et~al.}(1999)He, Lindström, Hagfeldt, and
  Lindquist]{He.1999}
He,~J.; Lindström,~H.; Hagfeldt,~A.; Lindquist,~S.-E. Dye-Sensitized
  Nanostructured p-Type Nickel Oxide Film as a Photocathode for a Solar Cell.
  \emph{The Journal of Physical Chemistry B} \textbf{1999}, \emph{103},
  8940--8943\relax
\mciteBstWouldAddEndPuncttrue
\mciteSetBstMidEndSepPunct{\mcitedefaultmidpunct}
{\mcitedefaultendpunct}{\mcitedefaultseppunct}\relax
\EndOfBibitem
\bibitem[Irwin \latin{et~al.}(2008)Irwin, Buchholz, Hains, Chang, and
  Marks]{Irwin.2008}
Irwin,~M.~D.; Buchholz,~D.~B.; Hains,~A.~W.; Chang,~R. P.~H.; Marks,~T.~J.
  $p$-Type semiconducting nickel oxide as an efficiency-enhancing anode
  interfacial layer in polymer bulk-heterojunction solar cells.
  \emph{Proceedings of the National Academy of Sciences} \textbf{2008},
  \emph{105}, 2783--2787\relax
\mciteBstWouldAddEndPuncttrue
\mciteSetBstMidEndSepPunct{\mcitedefaultmidpunct}
{\mcitedefaultendpunct}{\mcitedefaultseppunct}\relax
\EndOfBibitem
\bibitem[Park \latin{et~al.}(2010)Park, Kim, Kang, Kim, and Kang]{Park.2010}
Park,~S.-Y.; Kim,~H.-R.; Kang,~Y.-J.; Kim,~D.-H.; Kang,~J.-W. Organic solar
  cells employing magnetron sputtered p-type nickel oxide thin film as the
  anode buffer layer. \emph{Solar Energy Materials and Solar Cells}
  \textbf{2010}, \emph{94}, 2332--2336\relax
\mciteBstWouldAddEndPuncttrue
\mciteSetBstMidEndSepPunct{\mcitedefaultmidpunct}
{\mcitedefaultendpunct}{\mcitedefaultseppunct}\relax
\EndOfBibitem
\bibitem[Jeng \latin{et~al.}(2014)Jeng, Chen, Chiang, Lin, Tsai, Chang, Guo,
  Chen, Wen, and Hsu]{Jeng2014}
Jeng,~J.-Y.; Chen,~K.-C.; Chiang,~T.-Y.; Lin,~P.-Y.; Tsai,~T.-D.; Chang,~Y.-C.;
  Guo,~T.-F.; Chen,~P.; Wen,~T.-C.; Hsu,~Y.-J. Nickel Oxide Electrode
  Interlayer in {CH}3NH3PbI3Perovskite/{PCBM} Planar-Heterojunction Hybrid
  Solar Cells. \emph{Advanced Materials} \textbf{2014}, \emph{26},
  4107--4113\relax
\mciteBstWouldAddEndPuncttrue
\mciteSetBstMidEndSepPunct{\mcitedefaultmidpunct}
{\mcitedefaultendpunct}{\mcitedefaultseppunct}\relax
\EndOfBibitem
\bibitem[Karsthof \latin{et~al.}(2019)Karsthof, Grundmann, Anton, and
  Kremer]{Karsthof2019}
Karsthof,~R.; Grundmann,~M.; Anton,~A.~M.; Kremer,~F. Polaronic interacceptor
  hopping transport in intrinsically doped nickel oxide. \emph{Physical Review
  B} \textbf{2019}, \emph{99}\relax
\mciteBstWouldAddEndPuncttrue
\mciteSetBstMidEndSepPunct{\mcitedefaultmidpunct}
{\mcitedefaultendpunct}{\mcitedefaultseppunct}\relax
\EndOfBibitem
\bibitem[Allen \latin{et~al.}(1954)Allen, Stephenson, Stanford, and
  Bernstein]{Allen1954}
Allen,~R.~G.; Stephenson,~T.~E.; Stanford,~C.~P.; Bernstein,~S. Slow Neutron
  Cross Sections of Gold, Silver, Indium, Nickel, and Nickel Oxide.
  \emph{Physical Review} \textbf{1954}, \emph{96}, 1297--1305\relax
\mciteBstWouldAddEndPuncttrue
\mciteSetBstMidEndSepPunct{\mcitedefaultmidpunct}
{\mcitedefaultendpunct}{\mcitedefaultseppunct}\relax
\EndOfBibitem
\bibitem[Rose(1955)]{Rose.1955}
Rose,~A. Space-Charge-Limited Currents in Solids. \emph{Physical Review}
  \textbf{1955}, \emph{97}, 1538--1544\relax
\mciteBstWouldAddEndPuncttrue
\mciteSetBstMidEndSepPunct{\mcitedefaultmidpunct}
{\mcitedefaultendpunct}{\mcitedefaultseppunct}\relax
\EndOfBibitem
\bibitem[Seo \latin{et~al.}(2005)Seo, Lee, Kim, Ahn, Park, Kim, Yoo, Byun,
  Hwang, Kim, \latin{et~al.} others]{Seo.2005}
Seo,~S.; Lee,~M.~J.; Kim,~D.~C.; Ahn,~S.~E.; Park,~B.-H.; Kim,~Y.~S.;
  Yoo,~I.~K.; Byun,~I.~S.; Hwang,~I.~R.; Kim,~S.~H., \latin{et~al.}  Electrode
  dependence of resistance switching in polycrystalline {NiO} films.
  \emph{Applied Physics Letters} \textbf{2005}, \emph{87}, 263507\relax
\mciteBstWouldAddEndPuncttrue
\mciteSetBstMidEndSepPunct{\mcitedefaultmidpunct}
{\mcitedefaultendpunct}{\mcitedefaultseppunct}\relax
\EndOfBibitem
\bibitem[Chang \latin{et~al.}(2006)Chang, Lu, Kuo, and Wang]{Chang.2006}
Chang,~H.-L.; Lu,~T.~C.; Kuo,~H.~C.; Wang,~S.~C. Effect of oxygen on
  characteristics of nickel oxide/indium tin oxide heterojunction diodes.
  \emph{Journal of Applied Physics} \textbf{2006}, \emph{100}, 124503\relax
\mciteBstWouldAddEndPuncttrue
\mciteSetBstMidEndSepPunct{\mcitedefaultmidpunct}
{\mcitedefaultendpunct}{\mcitedefaultseppunct}\relax
\EndOfBibitem
\bibitem[Thimsen \latin{et~al.}(2012)Thimsen, Martinson, Elam, and
  Pellin]{Thimsen2012}
Thimsen,~E.; Martinson,~A. B.~F.; Elam,~J.~W.; Pellin,~M.~J. Energy Levels,
  Electronic Properties, and Rectification in Ultrathin p-NiO Films Synthesized
  by Atomic Layer Deposition. \textbf{2012}, \emph{116}, 16830--16840\relax
\mciteBstWouldAddEndPuncttrue
\mciteSetBstMidEndSepPunct{\mcitedefaultmidpunct}
{\mcitedefaultendpunct}{\mcitedefaultseppunct}\relax
\EndOfBibitem
\bibitem[Kremer and Sch\"{o}nhals(2002)Kremer, and Sch\"{o}nhals]{Kremer}
Kremer,~F.; Sch\"{o}nhals,~A. \emph{Broadband Dielectric Spectroscopy};
  Springer Berlin Heidelberg, 2002\relax
\mciteBstWouldAddEndPuncttrue
\mciteSetBstMidEndSepPunct{\mcitedefaultmidpunct}
{\mcitedefaultendpunct}{\mcitedefaultseppunct}\relax
\EndOfBibitem
\bibitem[Mitoff(1961)]{Mitoff.1961}
Mitoff,~S.~P. Electrical Conductivity and Thermodynamic Equilibrium in Nickel
  Oxide. \emph{The Journal of Chemical Physics} \textbf{1961}, \emph{35},
  882--889\relax
\mciteBstWouldAddEndPuncttrue
\mciteSetBstMidEndSepPunct{\mcitedefaultmidpunct}
{\mcitedefaultendpunct}{\mcitedefaultseppunct}\relax
\EndOfBibitem
\bibitem[Osburn and Vest(1971)Osburn, and Vest]{Osburn.1971a}
Osburn,~C.; Vest,~R. Defect structure and electrical properties of
  {NiO}{\textemdash}I. High temperature. \emph{Journal of Physics and Chemistry
  of Solids} \textbf{1971}, \emph{32}, 1331--1342\relax
\mciteBstWouldAddEndPuncttrue
\mciteSetBstMidEndSepPunct{\mcitedefaultmidpunct}
{\mcitedefaultendpunct}{\mcitedefaultseppunct}\relax
\EndOfBibitem
\bibitem[Serghei \latin{et~al.}(2009)Serghei, Tress, Sangoro, and
  Kremer]{Serghei.2009}
Serghei,~A.; Tress,~M.; Sangoro,~J.~R.; Kremer,~F. Electrode polarization and
  charge transport at solid interfaces. \emph{Physical Review B} \textbf{2009},
  \emph{80}\relax
\mciteBstWouldAddEndPuncttrue
\mciteSetBstMidEndSepPunct{\mcitedefaultmidpunct}
{\mcitedefaultendpunct}{\mcitedefaultseppunct}\relax
\EndOfBibitem
\bibitem[Ishai \latin{et~al.}(2013)Ishai, Talary, Caduff, Levy, and
  Feldman]{Ishai.2013}
Ishai,~P.~B.; Talary,~M.~S.; Caduff,~A.; Levy,~E.; Feldman,~Y. Electrode
  polarization in dielectric measurements: a review. \emph{Measurement Science
  and Technology} \textbf{2013}, \emph{24}, 102001\relax
\mciteBstWouldAddEndPuncttrue
\mciteSetBstMidEndSepPunct{\mcitedefaultmidpunct}
{\mcitedefaultendpunct}{\mcitedefaultseppunct}\relax
\EndOfBibitem
\bibitem[Shim and Moore(1957)Shim, and Moore]{Shim1957}
Shim,~M.~T.; Moore,~W.~J. Diffusion of Nickel in Nickel Oxide. \emph{The
  Journal of Chemical Physics} \textbf{1957}, \emph{26}, 802--804\relax
\mciteBstWouldAddEndPuncttrue
\mciteSetBstMidEndSepPunct{\mcitedefaultmidpunct}
{\mcitedefaultendpunct}{\mcitedefaultseppunct}\relax
\EndOfBibitem
\bibitem[Atkinson and Taylor(1979)Atkinson, and Taylor]{Atkinson1979}
Atkinson,~A.; Taylor,~R.~I. The diffusion of Ni in the bulk and along
  dislocations in NiO single crystals. \emph{Philosophical Magazine A}
  \textbf{1979}, \emph{39}, 581--595\relax
\mciteBstWouldAddEndPuncttrue
\mciteSetBstMidEndSepPunct{\mcitedefaultmidpunct}
{\mcitedefaultendpunct}{\mcitedefaultseppunct}\relax
\EndOfBibitem
\end{mcitethebibliography}

\end{document}